\documentstyle[11pt,epic, eepic, amsfonts]{article} 
\def\cqfd{\par\nopagebreak\rightline{\vrule height 3pt width 5pt depth 2pt} \medbreak} 
\def\cqfd{\par\nopagebreak\rightline{\vrule height 10pt width 5pt depth 3pt} \medbreak} 
 
\date{} 
\newtheorem{theoreme}{\bf Theorem}[section] \newtheorem{proposit}[theoreme]{\bf  
Proposition} \newtheorem{coro}[theoreme]{\bf Corollary}  
\newtheorem{lemme}[theoreme]{\bf Lemma} \newtheorem{defin}[theoreme]{\bf Definition}  
\newtheorem{rema}[theoreme]{\bf Remark} \newtheorem{exemple}[theoreme]{\bf Example}  
\newtheorem{condit}{\bf Condition} 
\newtheorem{fait}{\bf Claim}

\def\thm#1\par{\medskip\par\noindent\begin{theoreme} \strut \sl #1 \end{theoreme}\par} 
\def\propo#1\par{\medskip\par\noindent\begin{proposit} \strut \sl #1 \end{proposit} 
\par} 
\def\cor#1\par{\medskip\par\noindent\begin{coro} \strut \sl #1 \end{coro}\par} 
\def\lm#1\par{\medskip\par\noindent\begin{lemme} \strut \sl #1 \end{lemme}\par} 
\def\defil#1\par{\medskip\par\noindent\begin{condit} \strut \sl #1 \end{condit}\par} 
\def\fct#1\par{\medskip\par\noindent\begin{fait} \strut \sl #1 \end{fait} \cqfd\par} 
\def\defi#1\par{\medskip\par\noindent{\begin{defin} \strut  \sl #1 \end{defin}}\par} 
\def\nota#1\par{\par\noindent\begin{notat} \nopagebreak  \strut #1  \end{notat}} 
\def\rem#1\par{\par\noindent\begin{rema} \nopagebreak \strut \rm #1   \end{rema}} 
\def\ex#1\par{\par\noindent\begin{exemple} \nopagebreak \strut #1  \end{exemple}} 
\def\cqfd{\par\nopagebreak\rightline{\vrule height 3pt width 5pt depth 2pt} \medbreak} 
\def\N{\mbox{I\hspace{-.15em}N}}

\def\R{\mbox{I\hspace{-.15em}R}}

\begin{document}
\title{ 
{\bf Strategical languages of infinite words}}
\author{M. Arfi, B. Ould M Lemine, C. Selmi 
\thanks{LITIS EA 4108, 
Universit\'e du Havre, UFR Sciences et Techniques, 25 rue P. Lebon, BP 540, 76058 Le Havre Cedex, France
and Universit\'e de Rouen, D\'epartement d'Informatique, Avenue de l'Universit\'e, BP 12, F-76801 St. Etienne de Rouvray Cedex, 
France
\{mustapha.arfi, blemine, carla.selmi\}@litislab.eu  
} 
}
\maketitle
{\bf Abstract:}
We deal in this paper with strategical languages of infinite words, that is those generated 
by a nondeterministic strategy in the sense of game theory. 
We first show the existence of a minimal strategy for such languages, for which we give an explicit expression. 
Then we characterize the family of strategical languages as that of  closed ones, in the topological space of infinite words. 
Finally, we give a definition of a Nash equilibrium for such languages, that we illustrate with a famous example.
\section*{Introduction}
Game theory \cite{vMM} is usually defined as a mathematical tool used to analyze strategical interaction, the game, between 
individuals which are called players. 
The games studied in this paper are supposed simultaneous, noncooperative, infinitely repeated and with a perfect knowledge 
of the previous moves. We will elucidate these ideas through a famous example.\\
In game theory, the distinction between the cooperative and noncoperative game is crucial. 
The Prisoner's Dilemma \cite{Nash51} is an interesting example to explain these notions. 
It is a game involving two players where each one has two possible actions : cooperate (c) or  defect (d). 
The game consists of simultaneous actions (called moves) of both players. It can be represented using the matrix :
\begin{center}
\begin{tabular}{|c|c|c|}
\hline
$\pi$ & c & d\\
\hline
c & (4,4) & (0, 5)\\
\hline
d & (5, 0) & (1, 1)\\
\hline
\end{tabular}
\end{center}
where each entry $e_{ij}$ is an ordered pair of real numbers.
The two players are referred to as the row player and the column player respectively. The actions of the first player
are identified with the rows of the matrix and those of the second one with the columns. 
If the row player chooses  action $i$ and the second action $j$, the components of the ordered pair $e_{ij}$ are 
the payoff received by the first and the second player respectively. 
It is clear that if they could play cooperatively and make a binding agreement, they would both play c. 
If the game is noncooperative, the best action for each player is d.
\\
Suppose now that we consider infinite repetitions of a noncooperative base game.
This game is just as noncooperative as the base one, but it allows a certain form of interaction. 
Suppose that each player has a perfect knowledge of the previous moves of all the others. Then his strategy may depend 
on these previous moves and he may coordinate it with that of his opponents. 
For instance, if the base game is the Prisoner's Dilemma, grim-trigger is the strategy of cooperating in the first move 
and until your adversary defects, then of always defecting after the first defection of your opponent. 
Tit-for-tat is the strategy of playing at each step the action played by your adversary at the previous one ; 
the initial move is free.

In this paper, we make use of infinite words to analyze the kind of games we want to model. 
A match of such a game is represented as an infinite word on the alphabet $A$ of moves. 
In this context, a strategy for player $i$ can be viewed as a relation from the set of finite words on $A$  
to that of the actions of this player. 
The whole strategy of the game is defined as the vector composed by using the  strategies of all players. 
We can associate to each strategy vector a language $L$ of infinite words on $A$, 
defined as the set of all matches that the players  may make if everyone follows the strategy he decided to apply. 

Nash equilibrium is one of the most important notions in game theory.
The whole strategy of the game is defined as the vector composed by using the  strategies of all players. 
Intuitively, a strategy vector is a Nash equilibrium if one player's departure from it while the others remain 
faithful to it results in punishment. 
The idea is that once the players start playing according to such a strategy vector, then they all have a 
good reason to stay with it.

More precisely, our study will be organized as follows. In Sections 1 and 2, we introduce some basic notions 
on words and game theory. In Section 3, we show that
the same language can be generated by several strategies. We call ``strategical'' a language which is given by at
least one strategy. Section 4 is devoted to the proof of the existence of a minimal strategy for a strategical language, 
for which we give an explicit description. The characterization of the family of strategical languages as that of closed sets 
in the topological space of infinite words is given in Section 5. Finally, in Section 6,  we give a definition
of a Nash equilibrium for strategical languages.
%
%%%%%%%%%%%%%%%%%%%%%%%%%%%%%%%%%%%%%%%%%%%%%%%%%%%%%%%%%%%%%%%%%%%%%%%%%%%%%%%%%%%%%%%%%%%%%%%%%%%%%%%%%%%%%%%%%%%%%%%%%%%%%%%%%%%%%%%%%
%
\section{Words} 
A {\it word} is a finite sequence of elements of an alphabet $A$.
We denote by $A^*$ the set of all words on $A$.
The {\it length} of a word $w \in A^*$, denoted by $|w|$, is the number of letters of $A$ composing $w$.
Let $a \in A$. The empty word $\epsilon$ is the only word  of length zero. 
We denote by $|w|_a$ the number of the occurences of $a$ in $w$.
Given two words $u,v\in A^*$, we say that $u$ is a {\it factor} of $v$ if we have $v \in A^*uA^*$ and that
$u$ is a {\it prefix} or a {\it left factor} of $v$ if $v \in uA^*$ 
\\
An {\it infinite word} on $A$ is an infinite sequence $h$ of elements of $A$, 
which we will write $h = h_0h_1 \cdots h_t \cdots$. 
We denote by $A^{\omega}$ the set of all infinite words on $A$. Given a word $w \in A^*$ and an infinite word $h \in A^{\omega}$, 
we say that $w$ is a {\it prefix} or a {\it left factor} of $h$ if there exists an infinite word 
${h^{\prime}} \in A^{\omega}$ such that $h = wh^{\prime}$.\\
If $L$ is a subset  of $A^{\omega}$,  we denote by $Pref_k(L)$ the set of all words
that are prefixes of length $k$ of words of $L$. 
We set $Pref_{\ge k}(L) = \bigcup_{i \ge k}Pref_i(L)$ and simply $Pref(L) = Pref_{\ge 0}(L)$.\\
Finally, the left quotient of $L$ by a finite word $w$ is the subset $w^{-1}L$ of $A^{\omega}$ defined by 
$w^{-1}L = \{ h \in A^{\omega} | wh \in L\}.$
\section{Mathematical model for games}
Noncooperative games, in which moves consist of simultaneous actions of $n$ players, 
can be represented by a collection of $n$ utility functions. The values of these functions 
define the expected amount paid to the players.
A game is  a tuple $G = (P, A, \pi)$ where :
\begin{itemize}
\item $P =\{1, \cdots, n\}$, $n \in \N$, is the set of players.
\item $A_i$ is the set of the actions for player $i$.
\item $A = A_1 \times \ldots \times A_n$ is the alphabet of the moves.
\item $\pi_i: A \longrightarrow \R.$ is the utility function for player $i$.
\item $\pi = (\pi_1, \ldots, \pi_n): A \longrightarrow {\R}^n$ is the utility vector.
\end{itemize} 
We consider in this paper the $\delta$-discounted infinitely repeated game of $G$, which we note by $G^{\omega}$. 
In such a game, we model a match $h$ as an infinite sequence of moves which can be represented by an infinite word 
on the alphabet of the moves 
$A$ :  $h = h_0h_1 \cdots h_t \cdots \in A^{\omega}$.
\\
The utility with discounting factor $\delta \in (0, 1)$ of a match $h$ for player $i$ is defined as :
$$\pi^{\delta}_i(h) = (1 - \delta)\sum_{k=0}^{\infty }\pi_i(h_k)\delta^k.$$
\begin{exemple}
As concerns the Prisoner's Dilemma game, we have $P = \{1, 2 \}$, $A_1 = A_2 = \{c, d\}$, $A = \{c, d\} \times \{c, d\}$
and the utility function is defined by the matrix given in the Introduction. 
The infinite word $h = (c,c)^ {\omega}$ is an example of a match in which the two players cooperate infinitely.
\end{exemple}
%
%%%%%%%%%%%%%%%%%%%%%%%%%%%%%%%%%%%%%%%%%%%%%%%%%%%%%%%%%%%%%%%%%%%%%%%%%%%%%%%%%%%%%%%%%%%%%%%%%%%%%%%%%%%%%%%%%%%%%%%%%%%%%%%%%%%%%%%%%%%%%%%%
%
\section{Strategies and languages}
A nondeterministic strategy $\sigma_i$ for $G^{\omega}$, is a relation from $A^*$ into $A_i$ that describes 
the behaviour of player $i$ during the game.
A strategy vector on $A$ is the relation $ \sigma = (\sigma_1, \ldots, \sigma_n): A^* \longrightarrow  A$
defined by :
$$(a_1, \ldots, a_n) \in  \sigma(w) \; \Longleftrightarrow \; 
a_i \in \sigma_i(w), \; \forall w \in A^*, \; \forall a_i \in A_i, \; 1 \le i \le n.$$
Let $\Sigma$ be the set of all strategy vectors on $A$.
We consider now the map $\gamma : \Sigma \longrightarrow {\mathcal P}(A^{\omega})$, 
where ${\mathcal P}(A^{\omega})$ denotes the set of all languages in $A^{\omega}$, which associates to each strategy
$\sigma \in \Sigma$, the language of infinite words $\gamma(\sigma)$ given by 
$$\gamma(\sigma) = \{h \in A^{\omega} | \;  h_0 \in \sigma(\epsilon) \; and \; h_{t+1} \in \sigma(h_0 \cdots h_t), \; \forall t \ge 0 \}.$$
We also write $\sigma \rightarrow L$ to mean that $L = \gamma(\sigma)$.
\begin{exemple} 
We give a strategy for the Prisoner's Dilemma game.
$$
\sigma(w) = 
\left\{
\begin{array}{rl}
\{(c,c), (c, d)\} & \mbox{if} \; w \in (c,c)^*\\
\{(d,c), (d,d)\}  & \mbox{if} \; w \in (c,c)^*(c,d)((d,c) + (d,d))^*\\
\emptyset & \mbox{otherwise}
\end{array}
\right.
$$
It is usually called the "grim-trigger" strategy. The language $\gamma(\sigma)$ is described by the $\omega$-rational expression
$$(c,c)^{\omega} + (c,c)^{*}(c,d)((d,c) + (d,d))^{\omega}.$$
\end{exemple}
\begin{exemple} 
Consider the following strategy $\sigma$ on the alphabet $A = \{a, b\}$ :
$$
\sigma(w) = 
\left\{
\begin{array}{rl}
\{a, b\}  & \mbox{if} \; \mid w \mid_a < \mid w \mid_b\\
b  & \mbox{otherwise}
\end{array}
\right.
$$
The language $\gamma(\sigma)$ associated is 
$$\{h \in A^{\omega} \mid Pref(h) \in \{w \in A^* \mid |w|_a \leq |w|_b\}\}.$$
We note that this language is not $\omega$-rational, in the sense of language theory.
\end{exemple}
\begin{proposit}
The application $\gamma$ is neither injective, nor surjective, when $\mid A \mid >1$.
\end{proposit}
{\it \underline{Proof} :} Let $A = A_1 \times \ldots \times A_n$ be the alphabet of the moves and let $1 \le i \le n$ such that $\mid A_i\mid \ge 2$. Let $a_j \in A_j$, $1 \le j \le n$, and let $b_i \in A_i$, $b_i \neq a_i$.
\begin{enumerate}
\item 
$\gamma$ is not injective. Consider the strategies $\sigma$ and $\sigma^{\prime}$ defined as:
$$
\sigma(w) = 
\left\{
\begin{array}{rl}
(a_1, \ldots, a_n) & \mbox{if} \; w \in (a_1, \ldots, a_n)^*\\
\emptyset & \mbox{otherwise}
\end{array}
\right.
$$
and
$$
\sigma^{\prime}(w) = 
\left\{
\begin{array}{rl}
(a_1, \ldots, a_n) & \mbox{if} \; w \in (a_1, \ldots, a_n)^*\\
A & \mbox{otherwise}
\end{array}
\right.
$$
Obviously $\sigma$ and $\sigma^{\prime}(w)$ lead to the same language.
\item 
$\gamma$ is not surjective. Consider the language 
$$L = (a_1, \ldots, a_i, \ldots  a_n)^*(a_1, \ldots, b_i, \ldots  a_n)^{\omega}, a_i \neq b_i$$ 
We claim that there is no strategy $\sigma \in \Sigma$
verifying $\sigma \rightarrow L$. Indeed, suppose such a strategy exists. We then have
necessarily $ (a_1, \ldots, a_i, \ldots  a_n) \in \sigma((a_1, \ldots, a_i, \ldots  a_n)^t) \; \forall t \ge 0$, as a consequence 
of the expression of $L$.
Recall the definition of $L$ given at the beginning :
$$L = \{h \in A^{\omega} | h_0 \in \sigma(\epsilon), \; h_{t+1} \in \sigma(h_0 \cdots h_t) \; \forall t \ge 0 \}.$$
It implies that $(a_1, \ldots, a_i, \ldots  a_n)^{\omega} \in L$, because this word satisfies the required conditions, 
which leads us to a contradiction.
\end{enumerate}
\cqfd
For a language $L$, we note $S(L) = \{\sigma \in \Sigma | \; \sigma \longrightarrow L\}$, the set of strategies
generating $L$.
\begin{defin}
A language $L$ is strategical if there exists a strategy $\sigma \in \Sigma$ such that 
$\sigma \longrightarrow L$, that is if $S(L) \neq \emptyset$.
\end{defin}
%
%%%%%%%%%%%%%%%%%%%%%%%%%%%%%%%%%%%%%%%%%%%%%%%%%%%%%%%%%%%%%%%%%%%%%%%%%%%%%%%%%%%%%%%%%%%%%%%%%%%%%%%%%%%%%%%%%%%%%%%%%%%%%%%%%%%%%%%%
%
\section{Minimal strategy}
We define on the set $\Sigma$ of strategies on the alphabet $A$ the order relation given by :
$$\sigma \le \sigma^{\prime} \; \Leftrightarrow \; \sigma(w) \subset \sigma^{\prime}(w), \; \forall w \in A^*.$$
It is obviuos to remark that for every family $(\sigma_i)_{i \in I}$ of strategies of $\Sigma$,
\begin{itemize}
\item $\bigcap_{i \in I}\sigma_i \in \Sigma$;
\item $\bigcap_{i \in I}\sigma_i \subset \sigma_j, \; \forall j \in I.$
\end{itemize}
Among all the strategies giving a strategical language $L$, we consider a particular one :
$$\sigma_L = \bigcap_{\sigma \in S(L)}\sigma.$$
We have the following result.
\begin{proposit}
If $L$ is a strategical language, then $\sigma_L$ is the smallest strategy giving $L$. 
It will be called the minimal strategy of $L$.
\end{proposit}
{\it \underline{Proof} :} It suffices to show that $\gamma(\sigma_L) = L$. It will then be obvious that $\sigma_L$ 
is the smallest one. We obtain successively :
\begin{center}
\begin{tabular}{rll}
$\gamma({\sigma_L})$ & = & $\{h \in A^{\omega} | \; h_0 \in \sigma_L(\epsilon) \; and \; h_{t+1} \in \sigma_L(h_0 \ldots h_t), \forall t \ge 0 \}$\\
& = & $\{h \in A^{\omega} | \; h_0 \in \bigcap_{\sigma \in S(L)}\sigma(\epsilon) \; and \;
h_{t+1} \in \bigcap_{\sigma \in S(L)}\sigma(h_0 \ldots h_t), \; \forall t \ge 0\}$\\
& = & $\{h \in A^{\omega} | h_0 \in \sigma(\epsilon) \; and \;  h_{t+1} \in \sigma(h_0 \ldots h_t), \; 
\forall t \ge 0, \forall \sigma \in S(L)\}$\\
& = & $\bigcap_{\sigma \in S(L)}\{h \in A^{\omega} | h_0 \in \sigma(\epsilon) \; and \;  h_{t+1} \in \sigma(h_0 \ldots h_t), \;  \forall t \ge 0 \}$\\
& = & $\bigcap_{\sigma \in S(L)}\gamma(\sigma)$\\
& = & $\bigcap_{\sigma \in S(L)}L$\\
& = & $L$.
\end{tabular}
\end{center}
\cqfd
Consider now the particular strategy ${\hat{\sigma}}_L$ defined as follows :
\begin{center}
\begin{tabular}{cccl}
$ {\hat{\sigma}}_L$ :&  $A^*$ & $\longrightarrow$ & $A$ \\
& w & $\longmapsto$ & $Pref_1(w^{-1}L)$.\\
\end{tabular}
\end{center}
\begin{proposit} \label{Prop42}
For every  language $L \subset A^{\omega}$, we have $L \subset \gamma(\hat{\sigma}_L)$.
\end{proposit}
{\it \underline{Proof} :} Let $h = h_0 \ldots h_t \ldots \in L$. We have succesively :\\
$h_0 \in Pref_1(L) = Pref_1(\epsilon^{-1}L) = \hat{\sigma}_L(\epsilon)$,\\ 
$h_{t+1} \in Pref_1((h_0 \ldots h_t)^{-1}h) \subset Pref_1((h_0 \ldots h_t)^{-1}L), \; \forall t \ge 0$.\\ 
Thus, $h \in \gamma(\hat{\sigma}_L)$.
\cqfd
To characterize strategical languages of $A^{\omega}$, we introduce a new operator. For all subsets $X \subset A^*$, let
\begin{center}
$\overrightarrow{X} = \{u \in A^{\omega} | \; u \;$ has infinitely many prefixes in $X\}.$
\end{center}
\begin{exemple} We give the value of $\overrightarrow{X}$ for some simple sets $X \subset A^*$.
\begin{enumerate}
\item If $X = a^*b$ then $\overrightarrow{X} = \emptyset$.
\item If $X = (ab)^+$ then $\overrightarrow{X} = (ab)^{\omega}$.
\item If $X = (a+ b)^*b$ then $\overrightarrow{X} = (a^*b)^{\omega}$
\end{enumerate}
\end{exemple}
\begin{theoreme}
$L$ is strategical if and only if $L =\overrightarrow{Pref(L)}$.
\end{theoreme}
{\it \underline{Proof} :} We first prove that condition is necessary. The inclusion $L \subset \overrightarrow{Pref(L)}$ always holds.\\
Let now $h \in \overrightarrow{Pref(L)}$, then $h$ has infinitely many prefixes in $Pref(L) = \{w \in A^* \mid \exists x \in A^{\omega} : wx \in L\}$. This implies that: $\forall n \ge 0, \; \exists t_n \ge n, \; \exists x \in A^{\omega}$ such that $h_0 \ldots h_{t_n}x \in L$. 
We know that there exists a strategy $\sigma$ such that $L = \gamma(\sigma)$. 
Thus, $L = \{h \in A^{\omega} \mid h_0 \in \sigma(\epsilon) \; and \;  h_{t+1} \in \sigma(h_0 \ldots h_t), \; \forall t \ge 0\}$. 
Hence: $\forall n \ge 0, \; \exists t_n \ge n$
such that $h_0 \in \sigma(\epsilon)$ and $h_{t+1} \in \sigma(h_0 \ldots h_i) \; \forall i, 0 \le i \le t_n$. This implies
$h_0 \in \sigma(\epsilon)$ and $h_{i+1} \in \sigma(h_0 \ldots h_i), \; \forall i \ge 0$. Therefore $h \in L$.\\
We now prove the sufficient condition, that is $L = \overrightarrow{Pref(L)}$ then $\gamma(\hat{\sigma_L}) = L$. 
By Proposition \ref{Prop42}, we have only to establish the inclusion $\gamma(\hat{\sigma_L}) \subset L$. Let $h \in \gamma(\hat{\sigma_L})$. We have $h_0 \in \hat{\sigma_L}(\epsilon)$ and $h_{t+1} \in \hat{\sigma_L}((h_0 \ldots h_t)^{-1}L, \; \forall t \ge 0$.
That is, $h_0 \in Pref_1(L)$ and $h_{t+1} \in Pref_1((h_0 \ldots h_t)^{-1}L), \; \forall t \ge 0$. It is clear that $h_0 \in Pref(L)$ and $\forall t \ge 0, h_0 \ldots h_t \in Pref(L)$. It implies that $h$ admits an infinite number of left factors belonging to $Pref(L)$. Then $h \in L$ since $L = \overrightarrow{Pref(L)}$.
\cqfd
The proof that the condition of the previous proposition is sufficient implies the following result.
\begin{coro} \label{Cor45}
If $L$ is strategical, then $L = \gamma(\hat{\sigma_L})$.
\end{coro}
The results obtained so far can be summerized in this way :
\begin{proposit}
The following properties are equivalent :
\begin{itemize}
\item $L$ is a strategical language.
\item $\gamma(\hat{\sigma_L}) = L$.
\item $\overrightarrow{Pref(L)} = L$.
\end{itemize}
\end{proposit}
It is now possible to give an explicit description of the minimal strategy. In fact, for a strategical language $L$,
both strategies $\hat{\sigma_L}$ and  $\sigma_L$ coincide.
\begin{proposit}
For a strategical language $L \subset A^{\omega}$, we have $\hat{\sigma_L} = \sigma_L$.
\end{proposit}
{\it{\underline{Proof} :}} It suffices to show that $\hat{\sigma_L} \subset \sigma_L$. Let $w \in A^{\omega}$. 
We have $\hat{\sigma_L}(w) = Pref_1(w^{-1}L)$. Let $x \in \hat{\sigma_L}(w)$. 
Then $wx \in Pref(L)$. So : $\exists h \in L, \; \exists t \ge 0$ such that $w = h_0 \ldots h_t$ and $x = h_{t+1}$.
Since $h \in \gamma(\sigma_L)$, we have necessarily $h_{t+1} \in \sigma_L(h_0 \ldots h_t)$. Thus $x \in \sigma_L(w)$.
\cqfd
%
%%%%%%%%%%%%%%%%%%%%%%%%%%%%%%%%%%%%%%%%%%%%%%%%%%%%%%%%%%%%%%%%%%%%%%%%%%%%%%%%%%%%%%%%%%%%%%%%%%%%%%%%%%%%%%%%%%%%%%%%%%%%%%%%%%%%%%%%%%%%%%%%%%%%%
%
\section{Topological impact}
We consider on the set $A^{\omega}$ the distance $d$ defined as follows :
$$d(x, y) = (1+max\{|w| \mid w \in Pref(x) \cap Pref(y)\})^{-1}$$
with the convention $1/\infty = 0$.
\begin{proposit}
Equipped with this distance, $A^{\omega}$ is a complete metric space.
\end{proposit}
The next proposition is shown in [2,5].
%\cite{BoasNiv80, PePi04}.
\begin{proposit}
A language $L \subset A^{\omega}$ is closed if and only if $L = {\overrightarrow{Pref(L)}}$
\end{proposit}
\begin{coro}
$L$ is strategical if and only if $L$ is closed.
\end{coro}
It is usual to note $\overline L$ the smallest closed language containing $L$. Corollary \ref{Cor45} can be generalized 
to any language of $A^{\omega}$ in the following manner :
\begin{proposit}
For all language $L \subset A^{\omega}$, we have $\overline{L} = \gamma(\hat{\sigma_L})$.
\end{proposit}
{\it \underline{Proof} :} Since we always have $L \subset \gamma(\hat{\sigma_L})$, it is sufficient to prove 
that $\gamma(\hat{\sigma_L}) \subset {\overline{L}}$. Consider a word $h \in \gamma(\hat{\sigma_L})$. Recall that :\\
\begin{tabular}{rll}
$\gamma(\hat{\sigma_L})$ & = & $\{x \in A^{\omega} \mid x_0 \in \hat{\sigma_L}(\epsilon) \; and \;
x_{t+1} \in \hat{\sigma_L}(x_0 \cdots x_t), \; \forall t \ge 0\}$\\
& = & $\{x \in A^{\omega} \mid x_0 \in Pref_1(L) \; and \; x_{t+1} \in Pref_1((x_0 \cdots x_t)^{-1}L) \; \forall t \ge 0 \}$.
\end{tabular}
\\
We have then : 
$h_0 \in Pref_1(L)$ iff $\exists y_0 \in A^{\omega}, h_0 y_0 \in L$ and
$x_{t+1} \in Pref_1((x_0 \cdots x_t)^{-1}L)$ iff $\exists y_{t+1} \in A^{\omega}, h_0 \ldots h_{t+1} y_{t+1} \in L.$ It implies that : $\forall t \ge 0, \; \exists y_t \in A^{\omega}, \; h_0 \cdots h_ty_t \in L$. Define now the sequence $(h^{(t)}t)_{t \ge 0}$ of words of $L$ given by $h^{(t)} = h_0 \cdots h_ty_t$. It appears clearly that it admits the word $h$ as a limit. So $h \in {\overline L}$.
\cqfd
%
%%%%%%%%%%%%%%%%%%%%%%%%%%%%%%%%%%%%%%%%%%%%%%%%%%%%%%%%%%%%%%%%%%%%%%%%%%%%%%%%%%%%%%%%%%%%%%%%%%%%%%%%%%%%%%%%%%%%%%%%%%%%%%%%%%%%%%%
%
\section{Nash equilibrium}
Intuitively, a strategy vector is a Nash equilibrium if no player has any interest in leaving his strategy,
while his opponents remain faithful to theirs. 
Let us first introduce some basic notions, in order to give a formal definition of a Nash equilibrium.\\
Let $\alpha = (\alpha_1,  \ldots,  \alpha_n) \in A.$ We call {\it i-variation} of $\alpha$ 
every $\beta \in A$ such that $\alpha_i \neq \beta_i$ and $\alpha_j = \beta_j, \forall j \neq i$.
\\
Let $X$ be a language of $A^{\omega}.$ 
We call {\it i-variation} of a match $h = h_0h_1 \ldots h_t \ldots$ in $X$ every match 
$\overline h \in X$ for which there exists $t \geq 0$, an $i$-variation $\alpha$ of $h_t$ and a word
$w \in  A^{\omega}$ such that $\overline h = h_0 \ldots h_{t-1}{\alpha}w \in X$.
\\
A {\it good match } for player $i$ in $X$ is a match  $h \in X$ verifying 
$\pi_i(h) \geq \pi_i(\overline h)$ for every i-variation $\overline h$ of $h$.
Denote by $GM_i(X)$ the set of all good matches for player $i$ in $X$.
\begin{exemple}
Consider the language $L = (c,c)^{\omega} + (d,d)^{\omega}$. It is obvious that the words $(c,c)^{\omega}$ and
$(d,d)^{\omega}$ do not admit any $i$-variation in $L$. So we have $GM_i(L) = L, \; \forall i = 1, 2$.
\end{exemple}
\begin{exemple}
Let $L = (d,d)^{\omega} + (d,d)^*(d,c)((c,c) + (c,d))^{\omega}$.
This language involves the Prisoner's dilemma game strategy in which the first player defects as far as his adversary defects 
and cooperates infinitely as soon as his opponent cooperates.
We claim that $h = (d,c)(c,d)^{\omega} \in GM_2(L)$ if $\delta > 1/5$, otherwise $(d,d)^{\omega} \in GM_2(L)$. 
Indeed, let $\overline{h}$ be a $2$-variation of $h$.
Then $\overline{h} \in (d,d)^{\omega} + (d,d)^*((c,c) + (c,d))^{\omega}$. But, at the sight of the payment matrix
given in the Introduction, we notice it pay more for the second player always to choose d instead of c after his first cooperation. 
Thus, we will only examine the $2$-variations belonging to $(d,d)^{\omega} + (d,d)^*(d,c)(c,d)^{\omega}$.
We obtain successively for $n \in \N$ :
\begin{center}
\begin{tabular}{rll}
$\pi_2^{\delta}((d,d)^n(d,c)(c,d)^{\omega})$ &=& $(1-\delta)(\sum_{k=0}^{n}\delta^k + \sum_{k=n+2}^{\infty}5\delta^k)$\\
&=& $(1-\delta)[\sum_{k=0}^{n}\delta^k + 5(\sum_{k=0}^{\infty}\delta^k - \sum_{k=0}^{n+1}\delta^k)]$\\
%&=& ${(1-\delta^{n+1}) + 5[1 - (1-\delta^{n+2})]}$\\
&=& $1-\delta^{n+1} + 5\delta^{n+2}$\\
&=& $1 + \delta^{n+1}(5\delta-1)$.
\end{tabular}
\end{center}
The case of $\overline{h} = (d,d)^{\omega}$ can be dropped when $\delta > 1/5$, because we have
$1+\delta^{n+1}(5\delta-1) > 1 = \pi_2^{\delta}((d,d)^{\omega}).$
Furthemore, one can easily verify that the maximum of the function $n \longmapsto 1+\delta^{n+1}(5\delta-1)$ is reached for $n=0$.
Hence, the word $(d,c)(c,d)^{\omega}$ belongs to $GM_2(L)$.
\end{exemple}
The notion of Nash equilibrium also requires the introduction of some basic strategy vectors.
Let $\sigma = (\sigma_1, \ldots, \sigma_n)$ be a strategy vector and let $X = \gamma(\sigma)$ be the associated language. 
We denote by $\varpi_i$ :  $A^* \longrightarrow A_i$ the unpredictable strategy for player $i$, given by 
$\varpi_i(w) = A_i, \; \forall w \in A^*$.\\
We define for all $1 \leq i \leq n$, the following strategy vectors :
\begin{center}
\begin{tabular}{rll}
$\mu^{(i)}$ & = & $(\varpi_1, \ldots, \varpi_{i-1}, \sigma_i, \varpi_{i+1}, \ldots, \varpi_n)$\\
$\nu^{(i)}$ & = & $(\sigma_1, \ldots, \sigma_{i-1}, \varpi_i, \sigma_{i+1}, \ldots, \sigma_n).$
\end{tabular}
\end{center}
We denote by 
$X_i$ the language $\gamma(\mu^{(i)})$
and by 
$Y_i$ the language $\gamma(\nu^{(i)})$. 
\begin{proposit}
We have :
\begin{itemize}
\item $X  = \bigcap_{1 \leq i \leq n} X_i$;
\item $Y_i = \bigcap_{j \neq i}X_j \; \forall 1 \leq i \leq n.$
\end{itemize}
\end{proposit}
{\underline{\it Proof.}} For the first part, we obtain the succession of equations :
\begin{center}
\begin{tabular}{rll}
$X$ & = &  $\{h \in A^{\omega} | \; h_0 \in \sigma(\epsilon) \; and \; h_{t+1} \in \sigma(h_0 \ldots h_t), \; \forall t \ge 0 \}$\\
& = & $\{h \in A^{\omega} | \; h_{0,i} \in \sigma_i(\epsilon) \; and \; h_{t+1,i} \in \sigma_i(h_0 \ldots h_t),\; \forall 1 \leq i \leq n, \; \forall t \ge 0\}$\\
& = & $\bigcap_{1 \leq i \leq n}\{h \in A^{\omega} | \; h_{0,i} \in \sigma_i(\epsilon) \; and \; h_{t+1,i} \in \sigma_i(h_0 \ldots h_t), \; \forall t \ge 0\}$\\
& = & $\bigcap_{1 \leq i \leq n}\gamma(\mu^{(i)})$\\
& = & $\bigcap_{1 \leq i \leq n}X_i.$\\
\end{tabular}
\end{center}
The second part of the proof is easier, since we immediately obtain by using the lines above :
\begin{center}
\begin{tabular}{rll}
$Y_i$ & = & $\{h \in A^{\omega} | \; h_{0,j} \in \sigma_j(\epsilon) \; and \; h_{t+1,j} \in \sigma_j(h_0 \ldots h_t), \; \forall t \ge 0, \; \forall j \neq i\}$\\
& = & $\bigcap_{j \neq i}\gamma(\mu^{(j)})$\\
& = & $\bigcap_{j \neq i}X_j.$\\
\end{tabular}
\end{center}
\cqfd
\begin{defin}
A strategy vector $\sigma = (\sigma_1, \ldots, \sigma_n)$ is a Nash equilibrium if
$$\bigcap_{i=1}^{n}GM_i(Y_i) \neq \emptyset.$$
\end{defin}
In other words, a strategy vector is a Nash equilibrium if there exists a match that represents a good match 
for each player in the set of matches of the others \cite{Nash51}.
In particular, in the case of two players, the general definition becomes :
$$GM_1(X_2) \bigcap GM_2(X_1) \neq \emptyset.$$
\begin{exemple}
We consider in the Prisoner's Dilemma game, the vector $\sigma = (\sigma_1, \sigma_2)$ in which both players follow 
the grim-trigger strategy. In this case we have :
\begin{center}
\begin{tabular}{rll}
$X_1$ &=& $(c,c)^{\omega} + (c,c)^*(c,d)((c,c), (d,d))^{\omega},$\\
$X_2$ &=& $(c,c)^{\omega} + (c,c)^*(d,c)((c,d), (d,d))^{\omega}.$
\end{tabular}
\end{center}
We claim that
$(\sigma_1, \sigma_2)$ is a Nash equilibrium if and only if the discounting factor $\delta \geq 1/4$.
\\
Indeed $(c,c)^{\omega} \in GM_1(X_2) \cap GM_2(X_1)$.
Suppose $\overline{h} = (c, c)^{k-1}(c, d)(d, d)^{\omega}$ be a match with defection of the first player
at the rank $k \geq 0$.
We obtain after computations 
$$\pi_1^{\delta}(h) - \pi_1^{\delta}(\overline{h}) = \delta^{k}(4\delta - 1).$$
Then $\pi_1^{\delta}(h) - \pi_1^{\delta}(\overline{h}) \leq$ iff $\delta \leq 1/4$, which proves that
$h \in GM_1(X_2)$ whenever $\delta \geq 1/4$. In the same way, we can show that $h \in GM_2(X_1)$.
\end{exemple}
%%%%%%%%%%%%%%%%%%%%%%%%%%%%%%%%%%%%%%%%%%%%%%%%%%%%%%%%%%%%%%%%%%%%%%%%%%%%%%%%%%%%%%%%%%%%%%%%%%%%%%%%%%%%%%%%%%%%%%%%%%%%%%%%%
\section*{Conclusion and perspectives}
This paper was essentially devoted to a topological characterization of the family of strategical languages. It also
embeds a new definition of a Nash equilibrium that uses infinite words. 

Our purpose in the future will consist of analyzing more precisely the structure of good matches sets, then the
strategy vectors that admit Nash equilibria. We also wish to study mixed strategies (with probabilities).
It is obvious that this kind of strategies involves the nondeterministic ones.

%%%%%%%%%%%%%%%%%%%%%%%%%%%%%%%%%    bibliographie      %%%%%%%%%%%%%%%%%%%%%%%%%%%%%%%%%%%%%%%%%%%%%%%%%%%%%%%%%%%%%%%%%%%%%%%%%%%%

\bibliographystyle{style}

%\bibliography{/home/autil/latex/mabibli}

\end{document}